\documentclass[superscriptaddress, preprint, showkeys]{revtex4-1} 
\usepackage[pdftex]{graphicx}
\usepackage{dcolumn}
\usepackage{bm}

\usepackage[utf8]{inputenc}
\usepackage[T1]{fontenc}
\usepackage{mathptmx}
\usepackage[pdftex]{color}

\usepackage{bm}
\usepackage{ascmac}
\usepackage{float}
\usepackage{amsmath,amssymb,bm}
\usepackage{indentfirst} 
\usepackage[version=3]{mhchem}

\begin{document}

\title{
Electrical Manipulation of Magnetic Domain Structure 
in van der Waals Ferromagnetic \ce{Fe3GaTe2} 
Using Ferroelectric PMN-PT Single Crystal
}

\author{Riku Iimori}
\affiliation{Department of Physics, Kyushu University,
744 Motooka, Fukuoka, 819-0395, Japan}

\author{Yuta Kodani}
\affiliation{Department of Physics, Kyushu University,
744 Motooka, Fukuoka, 819-0395, Japan}

\author{Shaojie Hu}
\affiliation{Department of Physics, Kyushu University,
744 Motooka, Fukuoka, 819-0395, Japan}
\affiliation{College of Integrated Circuits and Optoelectronic Chips, 
Shenzhen Technology University, 
3002 Lantian Road, Pingshan District, 
Shenzhen Guangdong, China, 518118}

\author{Takashi Kimura}
\email[]{Corresponding author: t-kimu@phys.kyushu-u.ac.jp}
\affiliation{Department of Physics, Kyushu University,
744 Motooka, Fukuoka, 819-0395, Japan}
\affiliation{Research Center for Semiconductor and Device Ecosystem, 
Kyushu University, 6-1 Kasugakoen, Kasuga, 816-8580, Japan.}

\date{\today}
\begin{abstract}
2D van der Waals (vdW) ferromagnets have emerged 
as promising materials for spintronic applications due to their unique magnetic properties 
and tunability. 
Controlling ferromagnetism via external stimuli 
is critical for both fundamental research and device integration. 
In particular, modulation of magnetic anisotropy and exchange interactions 
through strain offers a viable pathway for functional control. 
Owing to their weak interlayer coupling, vdW ferromagnets exhibit pronounced sensitivity to
strain, enabling effective tuning of their magnetic states. 
In this study, electric-field-induced magnetoelectric coupling is investigated 
in the above-room-temperature vdW ferromagnet \ce{Fe3GaTe2} integrated 
on a ferroelectric PMN-PT substrate. 
It is demonstrated that application of an electric field leads to a substantial reduction 
in coercive force along with dynamic reconfiguration of the magnetic domain structure. 
These effects are attributed to electric-field-induced modulation of the vdW interlayer gap and enhancement of the Dzyaloshinskii-Moriya interaction. 
These findings reveal a strong interplay between electric fields and magnetism in vdW systems, offering a viable route toward the development of low-power, multifunctional magnetic devices. This work establishes a foundation for the electric-field control of magnetic properties 
in vdW ferromagnets and highlights their potential in next-generation spintronic technologies.
\end{abstract}

\maketitle

\section{INTRODUCTION}
2D van der Waals (vdW) ferromagnets have garnered significant attention due to their potential to host novel magnetic phenomena and their applicability in low-dimensional spintronic 
systems~\cite{vdW_appl_2017, vdW_appl_2017_2, vdW_appl_2019, vdW_appl_2019_2, vdW_appl_2021, vdW_appl_2021_2, vdW_appl_2021_3, vdW_appl_2021_4}. 
The distinctive crystal structure of these materials, composed of atomic layers with chemically inert surfaces held together by weak vdW interactions, provides an unparalleled platform for exploring the interplay between 2D material properties and magnetism~\cite{vdW_appl_2017, vdW_2Dmagnetism_2017, vdW_2Dmagnetism_2018, vdW_2Dmagnetism_2018_2, vdW_2Dmagnetism_2019, vdW_2Dmagnetism_2021}. 
This structural configuration enables the coexistence of low dimensionality and magnetic order, leading to intriguing phenomena such as flat-band magnetism~\cite{flatBand2022, flatBand2023}, the tunable quantum anomalous Hall effect~\cite{QAHE2015, QAHE2020, QAHE2021, QAHE2022, QAHE2023}, 
and room-temperature magnetic skyrmions~\cite{skyrmion2023, FGaTskyrmion2024, FGaTskyrmion2024_2, FGaTskyrmion2024_3, FGaTskyrmion2024_4}. 
From a technological perspective, vdW ferromagnets are particularly attractive because of their compatibility with the fabrication of atomic-layer devices via mechanical exfoliation. Their ultra-thin and mechanically robust nature makes them ideal candidates for flexible and stretchable electronic components, including wearable devices~\cite{Wearable2018, Wearable2022}. These properties also position vdW ferromagnets as promising materials for micro-electro-mechanical systems (MEMS) sensors, enabling novel operational mechanisms.

Despite these advantages, the stability of ferromagnetic order in 2D vdW materials remains a significant challenge. Although some materials exhibit Curie temperatures ($T_{\rm C}$) near room temperature~\cite{NearRoomTemp2019, NearRoomTemp2020}, 
the inherently low magnetic moments and susceptibility to thermal fluctuations in 2D systems hinder their practical applicability. 
As predicted by the Mermin-Wagner theorem, isotropic 2D systems with short-range interactions cannot sustain long-range magnetic order~\cite{MW1966}. 
However, magnetic anisotropy can stabilize such order, making it a critical parameter in defining the ferromagnetic properties of 2D materials. Addressing these challenges requires the discovery and development of vdW materials with enhanced magnetic anisotropy and stronger ferromagnetic interactions.  One promising candidate is \ce{Fe3GaTe2}, a vdW ferromagnet that exhibits ferromagnetic order above room temperature ($T_{\rm C} \sim 365$ K) with pronounced perpendicular magnetic anisotropy~\cite{zhangAboveroomtemperatureStrongIntrinsic2022}. 
Additionally, \ce{Fe3GaTe2} demonstrates high electrical and thermal spin-conversion efficiencies~\cite{Shuhan-san2023}, positioning it as a key material for next-generation spintronic devices. Building on these findings, we investigated the impact of applied pressure on the magnetic properties 
of \ce{Fe3GaTe2} thin films~\cite{RikuPressureFGaT2024}. 
Our study revealed that pressure systematically reduces the vdW interlayer gap, thereby enhancing perpendicular magnetic anisotropy at room temperature under relatively low pressures (below 1 GPa). Such behavior is rarely observed in conventional three-dimensional magnetic materials.  These results underscore a strong coupling between the vdW gap and magnetic anisotropy, highlighting the exceptional pressure responsiveness of \ce{Fe3GaTe2}. This discovery provides critical insights into the tunability of magnetic properties in 2D vdW ferromagnets, paving the way for innovative applications in spintronics and flexible electronics.

Expanding upon the idea of pressure-modulated magnetism in \ce{Fe3GaTe2}, we explored the potential for significant modulation of magnetic properties induced by in-plane strain transmitted from a substrate. Specifically, we investigated the use of ferroelectric substrates, where strain can be dynamically controlled through the application of an electric field. Previous studies have extensively analyzed field-induced modulation of magnetic properties in magnetic-ferroelectric hybrid devices~\cite{multiferro2011, multiferro2012, multiferro2012_2, multiferro2015, multiferro2016, TaniyamaSenseiMaltiferro2024}. The ferroelectric piezoelectric effect enables strain-mediated magnetic modulation solely via electric fields, eliminating current losses and offering a promising approach for nonvolatile memory technologies based on novel operating principles.  Moreover, the anisotropic piezoelectric strain and strain gradients generated by ferroelectric substrates can introduce additional symmetry breaking in vdW ferromagnets, potentially giving rise to emergent magnetic phenomena. To leverage these effects, we combined the ferroelectric piezoelectric properties of the substrate with the pronounced magnetoelastic coupling observed in \ce{Fe3GaTe2}, targeting effective electric-field modulation of its magnetic behavior.

In this study, we employed the ferroelectric crystal \ce{0.7Pb(Mg1/3Nb2/3)O3}-\ce{0.3PbTiO3} (PMN-PT), selected for its excellent ferroelectric characteristics, including a Curie temperature ($T_{\rm C}$) of approximately $416$ K, a high piezoelectric charge coefficient ($d_{33} \sim 2000$ pC/N), and an electromechanical coupling factor ($k_{33} \sim 0.94$)~\cite{PMN-PT_1990, PMN-PT_2007}. 
As illustrated in Fig.~\ref{fig:fig1}(a), we integrated PMN-PT with \ce{Fe3GaTe2} to examine electric-field-induced effects on the magnetic properties of the vdW ferromagnet.  Our experimental results demonstrate that the magnetic domain structure in \ce{Fe3GaTe2} can be effectively manipulated at room temperature through the inverse piezoelectric effect of PMN-PT under an applied electric field. These findings provide critical insights into the magnetic behaviors at vdW ferromagnet/ferroelectric interfaces and underscore the potential for energy-efficient electric-field control of magnetic structures. This work highlights a pathway for developing advanced spintronic devices with reduced power consumption and enhanced functional tunability.

\newpage
\section{RESULTS AND DISCUSSION}
To investigate the magnetic properties of the \ce{Fe3GaTe2}/PMN-PT heterostructure, we performed anomalous Hall effect (AHE) measurements. 
Figure~\ref{fig:fig3}(c) shows the transverse resistance ($R_{xy}$) curves observed at various temperatures. 
For ferromagnetic materials, the Hall resistance can be expressed as~\cite{AHE_1930, AHE_1932, AHE_2010}:
\begin{equation}
R_{xy} =R_{\rm OHE}B_{z} + R_{\rm AHE}M_{z},
\end{equation}
where the first term represents the ordinary Hall effect (OHE), and the second term corresponds to the anomalous Hall effect (AHE). The OHE contribution was neglected in this study due to its negligible magnitude compared to the dominant AHE component. The $R_{xy}$ curve at room temperature ($T = 297$ K) exhibits a nearly perfect rectangular hysteresis loop, indicative of robust perpendicular magnetic anisotropy under the standard conditions. Notably, the \ce{Fe3GaTe2} film prepared on a PMN-PT substrate displayed a more pronounced fully magnetized state at zero magnetic field compared to a similar film deposited on a thermally oxidized silicon substrate (\ce{SiO2}/Si).
As the temperature increased beyond room temperature, the AHE curves revealed that the magnetization reversal process was accompanied by the formation of multi magnetic domains prior to achieving full saturation. 
Figure~\ref{fig:fig3}(d) illustrates the temperature dependence of the AHE resistance ($\Delta R_{\rm AHE}$). Here, $2 \Delta R_{\rm AHE}$ 
is defined as the difference in transverse resistance between magnetic fields $\mu_0H = +185$ mT and $\mu_0H = -185$ mT. 
Here, we performed a fitting of $R_{xy}$ vs. $T$ based on the relationship 
$R_{xy} \propto (1-T/T_{\rm C})^{\beta}$.
As a result, the Curie temperature $T_{\rm C}$ of the ferromagnetic phase 
was determined to be $\sim 338$ K, 
which is consistent with the values reported in previous studies~\cite{zhangAboveroomtemperatureStrongIntrinsic2022, Shuhan-san2023, RikuPressureFGaT2024}.

We further examined the effect of an applied electric field on the AHE behavior of the \ce{Fe3GaTe2}/PMN-PT heterostructure. 
Figures~\ref{fig:fig4}(a) and \ref{fig:fig4}(b) present the AHE curves for electric fields ranging from $-12$ kV/cm to $+12$ kV/cm. The results reveal that the rectangular hysteresis loop evolves into a magnetic hysteresis loop characterized by a multi-domain structure under the application of an electric field ($E_{\rm [001]}$). 
This transformation was dependent on the magnitude of $E_{\rm [001]}$ but showed minimal sensitivity to the polarity of the spontaneous electric dipole in the PMN-PT.

To further elucidate the magnetization process, we evaluated the domain nucleation field ($H_{\rm n}$) and the saturation field ($H_{\rm s}$), 
as shown in Figs.~\ref{fig:fig5}(a) and \ref{fig:fig5}(b). 
As previously described, the application of an electric field transitions the magnetization process from an almost coherent rotation to a reversal mechanism associated with the formation of multi-domains. Figures~\ref{fig:fig5}(a) and \ref{fig:fig5}(b) present the dependence of $H_{\rm n}$ and $H_{\rm s}$ on the electric field ($E_{\rm [001]}$), where $E_{\rm [001]}$ was initially swept from $+12$ kV/cm to $-12$ kV/cm and then returned to $+12$ kV/cm. $H_{\rm n}$ decreases monotonically with increasing electric field, exhibiting a reduction of approximately $70~\%$ under an applied field of $12$ kV/cm. Notably, $H_{\rm n}$ displays a marked decrease with increasing magnitude of $E_{\rm [001]}$, forming a butterfly-shaped profile consistent with strain curves observed in piezoelectric materials.
This behavior suggests that the electric-field effect on $H_{\rm n}$ in \ce{Fe3GaTe2} originates 
from strain induced by the inverse piezoelectric effect of the PMN-PT substrate. 
Referring to the strain curve of the PMN-PT substrate 
in Fig. \ref{fig:fig2}(d), compressive in-plane strain is generated upon the application of $E_{\rm [001]}$. 
This strain is transferred to the \ce{Fe3GaTe2} layer, 
causing expansion along the $c$-axis via the Poisson effect, 
particularly within the van der Waals (vdW) gap.  
Conversely, a slight increase in $H_{\rm s}$ was observed near zero electric field; 
however, this variation remained within the error margins. 
We also observed that the Curie temperature ($T_{\rm C}$) of \ce{Fe3GaTe2} is insensitive to electric fields ($E_{\rm [001]}$) up to $6$ kV/cm, 
as presented in Fig.~\ref{fig:fig3}(e). 
To avoid potential damage to the PMN-PT substrate at elevated temperatures, the field dependence of $T_{\rm C}$ was not evaluated beyond $6$ kV/cm. The absence of significant field dependence of $T_{\rm C}$ aligns with the minimal field sensitivity of $H_{\rm s}$.

To further investigate the origin of electric-field effects on the magnetic properties of the \ce{Fe3GaTe2}/PMN-PT heterostructure, we analyzed the difference $\Delta H$ between the saturation field $H_{\rm s}$ and the nucleation field $H_{\rm n}$ of the multi-domain structure. This difference characterizes the stabilization region of metastable domains. 
Figure~\ref{fig:fig5}(c) illustrates the stabilization field range of multi-domains as a function of the electric field. The results reveal that higher electric fields enhance the stability of the multi-domain structure, indicating the pivotal role of electric-field-induced strain in modulating the magnetic properties of vdW ferromagnetic systems.

To understand the electric-field dependence of $\Delta H$, 
we simply propose that the observed reduction in perpendicular magnetic anisotropy arises 
from an expansion of the van der Waals (vdW) gap. 
Our previous studies on the pressure effects in \ce{Fe3GaTe2} have demonstrated 
that applying the pressure significantly enhances perpendicular magnetic anisotropy~\cite{RikuPressureFGaT2024}.
According to the analysis, the pressure is known to induce the modification of the vdW gap as well as the lattice parameters effectively, namely a few percent per GPa.  On the other hand, as shown in Fig. \ref{fig:fig2}(d), the induced strain via the inverse piezo electric effect is one order smaller than that 
by the pressure application (see also Note S5, Supporting Information).  
This means that the observed enhancement $\Delta H$ due to the electric field 
cannot be explained only by the modification of the PMA due to the reduction in the vdW gap.

To further discuss the mechanisms underlying the observed effects more quantitatively, 
we take into account the Dzyaloshinskii-Moriya interaction (DMI).  
The magnetic energy density $w_{\rm e}$ for the present system can be expressed as:
\begin{equation}
    w_{\rm e} = -A(\nabla {\bf m})^2 -K_{\rm ani}m_z^2 + D{\bf m} \cdot (\nabla \times {\bf m}),
\end{equation}
where $A$ represents the exchange stiffness constant, $K_{\rm{ani}}$ is the magnetic anisotropy constant, 
and $D$ denotes the DMI constant.  
With regard to the ferromagnetic exchange interaction in \ce{Fe3GaTe2}, our observations reveal that the Curie temperature $T_{\rm C}$ remains unchanged up to $E_{\rm [001]} = 6$ kV/cm as indicated in Fig.~\ref{fig:fig3}(e). Additionally, as shown in Fig.~\ref{fig:fig5}(a), the saturation field ($H_{\rm s}$) displays minimal sensitivity to the applied electric field. Furthermore, the in-plane strain in the PMN-PT substrate induced by the electric field is less than 0.03\%. These results strongly support the hypothesis that the ferromagnetic exchange stiffness constant ($A$) within the 2D layers of \ce{Fe3GaTe2} is not significantly influenced by the application of the electric field.

Although the strain induced by the electric field is relatively small, it permeates the entire thickness 
of the Fe$_3$GaTe$_2$ layer. The vdW gap is notably sensitive to strain, which modulates 
the perpendicular magnetic anisotropy. However, as previously discussed, these strain-induced effects 
alone cannot fully account for the significant reduction in the nucleation field $H_n$ under the electric field application. Importantly, the strain decreases progressively with distance from the interface, creating a strain gradient that suggests the relevance of the DMI.  An enhancement in DMI can lead to the formation of non-collinear spin structures, such as skyrmions, in ferromagnetic materials. Notably, room-temperature skyrmions attributed to DMI have been reported in Fe$_3$GaTe$_2$~\cite{FGaTskyrmion2024, FGaTskyrmion2024_2, FGaTskyrmion2024_3, FGaTskyrmion2024_4}. DMI arises from the interplay of spatial inversion symmetry breaking and strong spin-orbit coupling. In the Fe$_3$GaTe$_2$/PMN-PT heterostructure studied here, strain gradients due to strain relaxation significantly break symmetry along the out-of-plane direction. Additionally, the anisotropic strain introduced by the single-crystal PMN-PT substrate, characterized by its four-fold symmetry, 
disrupts the intrinsic hexagonal symmetry of Fe$_3$GaTe$_2$. These combined factors contribute to the 
enhancement of DMI in Fe$_3$GaTe$_2$, as illustrated in Fig.~\ref{fig:fig5}(e).

To further investigate these phenomena, direct observations of magnetic domain behavior under electric field 
modulation would provide valuable insights. 
The ability to fabricate few-atomic-layer \ce{Fe3GaTe2} films directly on PMN-PT with clean interfaces would allow for maximized strain coupling and could unveil more pronounced or novel electric-field effects intrinsic to the ultrathin regime.
The interface multiferroic structure formed by Fe$_3$GaTe$_2$ 
on PMN-PT demonstrates the potential for controlling magnetic domain states with minimal electric power 
consumption. This capability offers promising applications in electric-field-driven micro-electromechanical 
systems (MEMS) and next-generation memory devices.

\section{CONCLUSIONS}
We experimentally investigated the electric-field effects on the magnetic properties 
of a heterostructure composed of the vdW ferromagnet \ce{Fe3GaTe2} and the ferroelectric PMN-PT. 
To evaluate the magnetic properties, AHE measurements were performed under varying back-gate electric fields and temperatures. 
Application of an electric field resulted in a significant reduction of the domain nucleation field and markedly altered the magnetization reversal mechanism from coherent rotation 
in a single-domain state to the formation of a multi-domain structure. In contrast, no notable change in the Curie temperature, associated with ferromagnetic exchange interactions,
was observed under electric field application. 
Our analysis revealed that the inverse piezoelectric
effect induced by the PMN-PT expanded the vdW gap, which not only weakened the interlayer
perpendicular magnetic anisotropy but also enhanced the DMI in \ce{Fe3GaTe2} through symmetry
reduction. 
These findings highlight the strong magneto-elastic coupling properties of vdW ferromagnets and demonstrate the potential for electric-field-driven control of magnetic structures
with minimal power consumption. This approach paves the way for advanced applications in
next-generation spintronic devices.

\clearpage

\newpage
\section*{Experimental Section}
\begin{figure}
    \begin{center}
    \includegraphics[width=90mm]{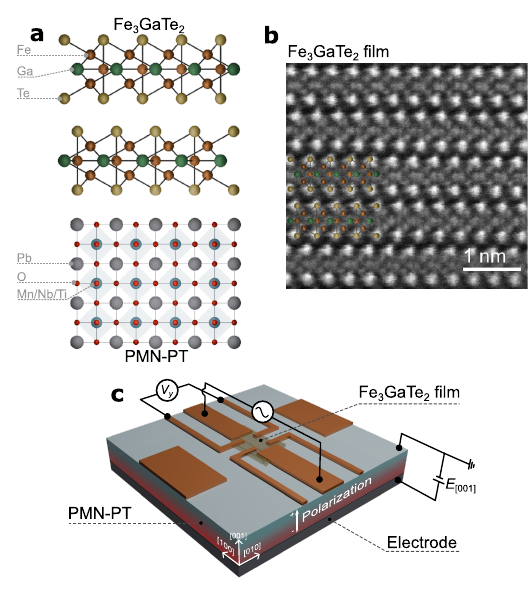}
    \end{center}
    \caption{
    {\bf Heterostructure combining vdW ferromagnetic \ce{Fe3GaTe2} and ferroelectric PMN-PT.}
    (a) Schematic of atomic structure of \ce{Fe3GaTe2}/PMN-PT heterostructure.
    Here, the composition of the PMN-PT is \ce{0.7Pb(Mg1/3Nb2/3)O3}-\ce{0.3PbTiO3}, 
    and a $c$-plane cut PMN-PT substrate was used in this study. 
    (b) Cross-sectional transmission electron microscopy (TEM) image of the \ce{Fe3GaTe2} film.
    (c) Device structure for the magneto-transport measurement with circuit diagram. The electric field is applied to the ferroelectric PMN-PT substrate 
    in a back-gate configuration. 
    The inverse piezoelectric effect of the substrate transmits the strain 
    to the \ce{Fe3GaTe2}.
    }
    \label{fig:fig1}
\end{figure}

\begin{figure}
    \begin{center}
    \includegraphics[width=165mm]{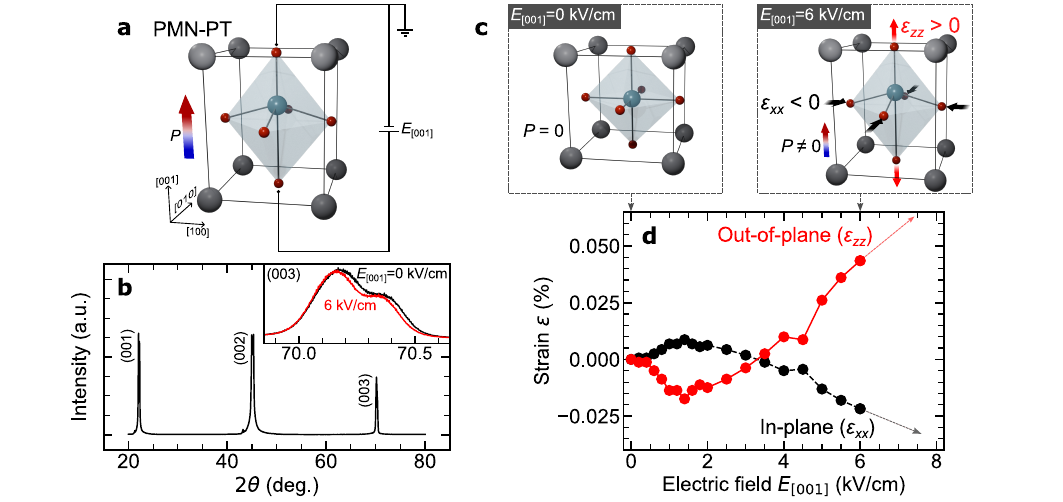}
    \end{center}
    \caption{
	{\bf Electric field characteristics of the PMN-PT single crystal.}
    (a) Schematic illustration of the inverse piezoelectric effect of the PMN-PT crystal under the electric field $E_{\rm [001]}$. In this study, the electric field $E_{\rm [001]}$ is applied along the [001] direction of PMN-PT single crystal.
    (b) X-ray diffraction (XRD) pattern of the PMN-PT single crystal. 
    The inset shows the (003) spectra as a function of 
    the back-gate electric field up to 6 kV/cm. 
    (c) Schematic diagrams of the polarization and strain profiles under $E_{[001]}=0$~kV/cm and 6~kV/cm.
    (d) Electric field dependence of out-of-plane strain ($\epsilon_{zz}=\Delta c/c$) 
    evaluated from the shift of the (003) XRD peak.
    In-plane ($\epsilon_{xx}$) strain was derived based on the approximation of constant volume 
    in the Poisson effect ($2\epsilon_{xx}+\epsilon_{zz}=0$).
    }
    \label{fig:fig2}
\end{figure}

\begin{figure}
    \begin{center}
    \includegraphics[width=165mm]{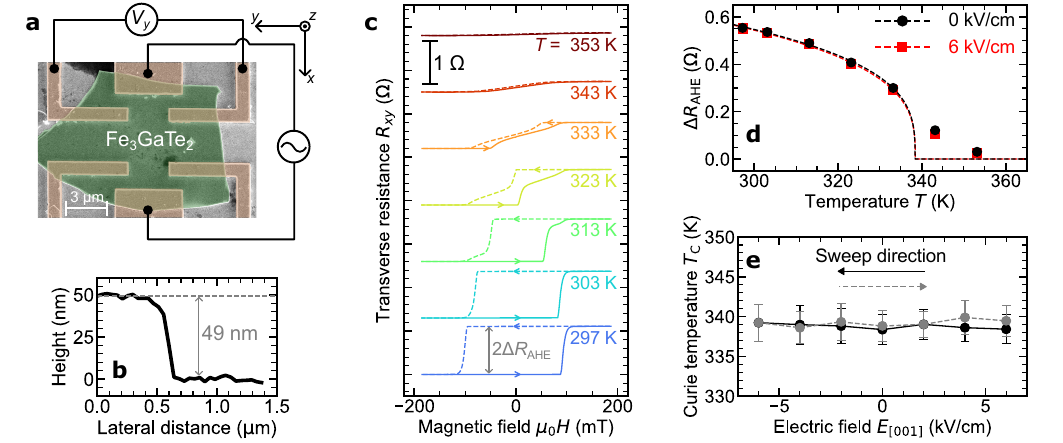}
    \end{center}
    \caption{
	{\bf Ferromagnetic properties of the \ce{Fe3GaTe2}/PMN-PT device.}
     (a) Scanning electron microscope (SEM) image of 
    the fabricated \ce{Fe3GaTe2}/PMN-PT device together with the probe configuration.
	(b) Cross-sectional height profile measured by the atomic force microscopy (AFM). 
	(c) Transverse resistance $R_{xy}$ as a function of 
    the applied magnetic field $H$ along the perpendicular direction 
    under various temperatures. 
	(d) Temperature dependence $T$ of the anomalous Hall resistance 
    $\Delta R_{\rm AHE}$ at $E_{\rm [001]}=0$ and 6 kV/cm. 
Curie temperature $T_{\rm C}$ of 
    the ferromagnetism for the \ce{Fe3GaTe2}/PMN-PT heterostructure 
    was determined by fitting the $\Delta R_{\rm AHE}$ vs. $T$ data to the relationship 
$\Delta R_{\rm AHE} = A (1 - T / T_{\rm C})^{\beta}$.
Here, $A$, $T_{\rm C}$ and $\beta$ are fitting parameters. 
The fitting was conducted for the temperature range $T \leq 333$~K, which is below the Curie temperature. This range was chosen to avoid the influence of non-monotonic changes in the anomalous Hall effect (AHE) near $T_{\mathrm{C}}$, which can arise from spin fluctuations and associated modifications in the electronic structure.
For example, at an applied back-gate electric field of $E_{\rm [001]} = 6$~kV/cm, the fitting parameters were obtained as $A = 1.06 \pm 0.10$~${\rm \Omega}$, 
$T_{\rm C} = 338.4 \pm 1.9$~K, and $\beta = 0.303 \pm 0.044$.
	(e) Curie temperature $T_{\rm C}$ as a function of 
the applied electric field $E_{\rm [001]}$.
    }
    \label{fig:fig3}
\end{figure}

\begin{figure}
    \begin{center}
    \includegraphics[width=90mm]{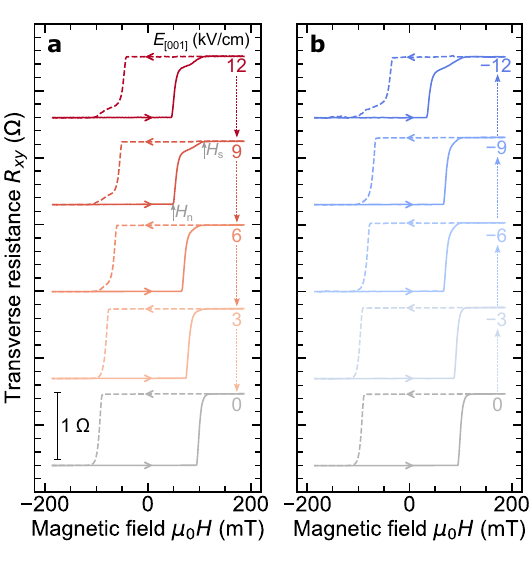}
    \end{center}
    \caption{
	{\bf Electric-field modulated anomalous Hall resistance curves in the \ce{Fe3GaTe2}/PMN-PT device.}
	Room-temperature anomalous Hall resistance $R_{xy}$ curves measured under different back-gate electric fields $E_{[001]}$ ranging (a) from 0 to $+12$ kV/cm (positive direction) and (b) from 0 to $-12$ kV/cm (negative direction).
    }
    \label{fig:fig4}
\end{figure}

\begin{figure}
    \begin{center}
    \includegraphics[width=165mm]{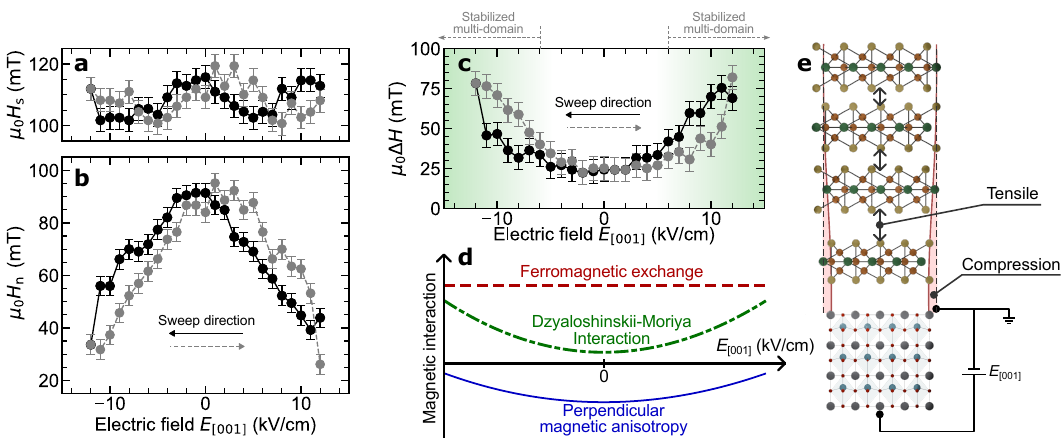}
    \end{center}
    \caption{
    {\bf Origin of the electric field effect on the magnetic domains of the \ce{Fe3GaTe2}/PMN-PT heterostructure.}
    (a) Electric field $E_{\rm [001]}$ dependence of the saturation magnetic field $H_{\rm s}$. 
    (b) The nucleation magnetic field $H_{\rm n}$ as a function of the $E_{\rm [001]}$.
    Here, the black solid line represents the magnetic field being swept from positive to negative, 
    while the gray dashed line indicates the opposite direction.
    (c) Electric field $E_{\rm [001]}$ dependence of the $\Delta H$ ($=H_{\rm s} - H_{\rm n}$) 
in the \ce{Fe3GaTe2}/PMN-PT heterostructure.
    (d) Changes in magnetic interactions induced by electric field $E_{\rm [001]}$.
Since there was no significant change in the Curie temperature, the ferromagnetic interactions remained largely unaffected. On the other hand, the in-plane strain of PMN-PT expands the van der Waals gap, leading to a reduction in perpendicular magnetic anisotropy. Additionally, the change in symmetry due to the strain gradient enhances the Dzyaloshinskii-Moriya interaction.
    (e) Strain profile induced by the electric field in 
    the \ce{Fe3GaTe2}/PMN-PT structure.
    }
    \label{fig:fig5}
\end{figure}

{\bf Device Fabrication and Characterization}\\
In the present study, we fabricated the \ce{Fe3GaTe2}/PMN-PT heterostructure device by mechanically exfoliating single crystals consisting of \ce{Fe3GaTe2}. The bulk \ce{Fe3GaTe2} crystals were synthesized using a self-flux method. A mixture with a molar ratio of Fe:Ga:Te = 1:1:2 was placed in an alumina crucible, vacuum-sealed in a quartz tube, and heat-treated in a muffle furnace. The temperature was ramped to 1000~$^\circ$C within 1 hour, maintained for 24 hours, then lowered to 880~$^\circ$C within 1 hour, and gradually cooled to 780~$^\circ$C over 120 hours. The temperature was held at 780~$^\circ$C for 24 hours before centrifugation was employed to separate the crystals from the Te flux.
Figure S1 (Supporting Information) presents the X-ray diffraction (XRD) spectra of the synthesized \ce{Fe3GaTe2}, which predominantly exhibit (00$h$) peaks, indicating a high-quality crystal with $c$-axis orientation.
In addition, the transmission electron microscopy (TEM) image clearly reveals the layered crystal structure as shown in Fig.~\ref{fig:fig1}(b).
Figure~\ref{fig:fig1}(c) illustrates the Hall measurement device structure. The \ce{Fe3GaTe2} film was mechanically exfoliated and transferred onto a (001)-oriented single-crystalline PMN-PT substrate in a nitrogen-filled glovebox. The thickness of the exfoliated \ce{Fe3GaTe2} film was measured to be 49~nm using atomic force microscopy (AFM), as shown in the right panel of Fig.~\ref{fig:fig3}(b).
Subsequently, electrodes were patterned using electron beam lithography and low-energy ion milling techniques. Figure~\ref{fig:fig3}(a) shows a scanning electron microscopy (SEM) image of the fabricated \ce{Fe3GaTe2}/PMN-PT device. Copper (200~nm) electrodes were deposited via Joule heating evaporation under a base pressure of $2.0 \times 10^{-6}$~Pa. For back-gating control, silver paste was applied to the back side of the PMN-PT substrate to fabricate gate electrodes. To prevent property changes due to redox reactions, the fabricated devices were spin-coated with poly(methyl methacrylate) (PMMA).

{\bf Electric Field Control Using Ferroelectric PMN-PT Single Crystals}\\
The electric field control was performed using the programmable DC voltage source Advantest R6161.
The electric field was symmetrically driven in a bipolar manner, 
alternating sequentially between $|E_{\rm max}|$ and $-|E_{\rm max}|$, to prevent the retention of biased hysteresis. 
Furthermore, to prevent cracks caused by inhomogeneous strain, the uniform application of the gate electric field was verified using an optical microscope by confirming that the transparency changed uniformly with 
the application of the electric field.
The detailed electric-field-induced strain characteristics of the PMN-PT were evaluated using the XRD 
as shown in the inset of Fig. 2(b).
As the strain-induced peak shift becomes more pronounced at higher diffraction angles, we consequently selected the (003) peak at the highest accessible angle for our system. In the measurement of electric-field-induced peak shifts, it was necessary to evaluate the XRD spectra using the highest resolution available with our setup. To ensure efficient measurements and to avoid time-dependent drift, the XRD scans were limited to the region around the (003) peak as shown in Note S2 (Supporting Information).
Figure \ref{fig:fig2}(d) shows the gate field $E_{[001]}$ dependence of the out-of-plane strain 
$\epsilon_{zz}$ ($=\Delta c/c$) 
for the PMN-PT substrate. The strain curve was evaluated by the Bragg-peak shift of the PMN-PT (003) reflection. 
We observed the ferroelectric hysteresis and the tensile strain of about $0.05 \%$ 
along the $c$-axis, which is consistent with the previous studies~\cite{PMN-PT_strainCurve}. 
Considering the constant volume Poisson effect ($2\epsilon_{xx}+\epsilon_{zz}=0$), 
the $c$-axis tensile strain corresponds to an in-plane compressive strain $\epsilon_{xx}$
of about $-0.03~\%$, suggesting that \ce{Fe3GaTe2} undergoes in-plane compressive strain 
from the PMN-PT interface as shown in Figs.~\ref{fig:fig1}(c) and~\ref{fig:fig2}(d).

{\bf Magneto-Electric Transport Properties Measurements}\\
The anomalous Hall effect (AHE) measurements were performed using an AC lock-in detection technique under temperature control from 300 to 400~K. The applied current was 100~$\mu$A at a frequency of 173~Hz. Here, the time constant of the lock-in amplifier is 100 ms and the sweep rate of the external magnetic field is 25 Oe/s.  All measurements were conducted in vacuum to prevent degradation of the device during experiments.
For high-temperature application from 300~K to 400~K, a ceramic heater (MC2525H) manufactured by Sakaguchi electric heaters Co., Ltd. was used, and temperature control was achieved using a DC-type temperature controller with a K-type thermocouple.

{\bf First-Principles Calculations}\\
To evaluate the effect of in-plane substrate strain on \ce{Fe3GaTe2}, 
first-principles calculations based on density functional theory (DFT) were performed using the QUANTUM ESPRESSO package~\cite{QE_2009, QE_2017}.
We employed pseudopotentials within the projector-augmented-wave (PAW) framework, using the Perdew–Burke–Ernzerhof (PBE) generalized gradient approximation (GGA) for the exchange-correlation functional. The plane wave cutoff energy was set to 70 Ry, 
with a $12\times12\times4$ $k$-point mesh for self-consistent calculations in the Brillouin zone. To account for van der Waals interactions, we applied Grimme's DFT-D3 dispersion correction~\cite{DFTD3_2006, DFTD3_2010}.
Based on the above, the lattice relaxation of bulk \ce{Fe3GaTe2} was performed under various in-plane strain conditions to evaluate the change in the lattice parameter along the $c$-axis. 
The detailed results are provided in Note S3 (Supporting Information).


\section*{Author Contributions}
R.I., H.S. and T.K. conceptualized and planned the project. 
T.K. supervised the project. 
R.I. fabricated samples.  
Y.K. evaluated the X-ray diffraction measurements with support from R.I.
R.I. and Y.K. performed transport measurements. 
Analyses of the data were done by R.I., Y.K., H.S., and T.K. 
R.I. wrote the manuscript with support from K.Y., H.S. and T.K. 
All authors discussed the results and commented on the manuscript.

\section*{Acknowledgments}
This work was supported by NEDO (Uncharted Territory Challenge 24000474), 
JSPS KAKENHI Grant Numbers 21H05021, 23KJ1701, 
and JST (CREST: JPMJCR18J1, SICORP: 22480474).

\section*{Conflict of Interest}
The authors declare no conflict of interest.

\section*{Supporting Information}
Supporting information is available for this paper at xxx.

\newpage

\end{document}